\documentclass{article}
\usepackage{preprint}
\usepackage{hyperref}
\usepackage{cite}
\usepackage{amsmath,amssymb,amsfonts}
\usepackage{algorithmic}
\usepackage{graphicx}
\usepackage{textcomp}
\usepackage{xcolor}
\usepackage{url}
\usepackage{cleveref}
\usepackage{tikz}
\usepackage{subcaption}

\newcommand{\name}{Kirin}
\newcommand*\circled[1]{
    \tikz[baseline=(char.base)]{
            \node[shape=circle,draw,inner sep=0.75pt] (char) {#1};
    }
    \hspace{-0.5em}
}
\newcommand\copyrighttext{%
  \footnotesize \textcopyright \the\year{} IEEE. Personal use of this material is permitted. Permission from IEEE must be obtained for all other uses.}

\usepackage{booktabs}
\usepackage{multirow}
\definecolor{lgreen}{RGB}{0.12,0.92,0.07}
\definecolor{lime}{HTML}{A6CE39}
\DeclareRobustCommand{\orcidicon}{
	\begin{tikzpicture}
	\draw[lime, fill=lime] (0,0)
	circle [radius=0.16]
	node[white] {{\fontfamily{qag}\selectfont \tiny ID}};
	\draw[white, fill=white] (-0.0625,0.095)
	circle [radius=0.007];
	\end{tikzpicture}
	\hspace{-2mm}
}

\def\BibTeX{{\rm B\kern-.05em{\sc i\kern-.025em b}\kern-.08em
    T\kern-.1667em\lower.7ex\hbox{E}\kern-.125emX}}

\usepackage{authblk}
\author{Joshua Bauer}
\author{Sebastian Werner~\href{https://orcid.org/0000-0001-8051-7226}{\orcidicon}}
\author{Maria C. Borges~\href{https://orcid.org/0000-0003-1661-5969}{\orcidicon}}
\affil{Technische Universität Berlin \\
\texttt{\{joba, sw, mb\}@ise.tu-berlin.de}}
\begin{document}

\title{\name{}: Cloud-native WebAssembly Service Orchestration}

\twocolumn[\begin{@twocolumnfalse}
\maketitle

\begin{abstract}
Modern cloud computing infrastructure relies heavily on virtualization to provide workload isolation and resource efficiency. While containers have become the dominant deployment primitive due to their fast orchestration compared to traditional virtual machines, they still introduce non-trivial overhead. WebAssembly (Wasm) has emerged as an alternative isolation technology, offering a lightweight execution model. Because of compatibility limitations, WebAssembly has mainly been applied to Function-as-a-Service (FaaS) and Edge Computing scenarios, where it has attracted growing research interest. 

However, recent advances in the WebAssembly ecosystem have significantly matured the technology, raising the question of whether it can serve as a viable replacement for containers in cloud-native workloads. In this paper, we present Kirin, an orchestrator for cloud-native WebAssembly services. It supports core orchestration responsibilities, including resource scheduling, lifecycle management, scaling, and request routing. We evaluate Kirin against container-based deployments for several service workloads and demonstrate promising results, particularly in service lifecycle management. Simultaneously, we confirm known limitations in CPU-intensive workloads. However, for typical service workloads, Kirin demonstrates competitive performance, suggesting that WebAssembly is a viable candidate for broader adoption as a cloud-native deployment technology.

\end{abstract}

\end{@twocolumnfalse}]

\section{Introduction}
Modern cloud-native applications are typically built as a set of loosely coupled services that operate independently of one another~\cite{kratzke_cloudnative_2017}.
Containers have become the de facto standard for deploying these services, offering several advantages such as process isolation, portability across environments, and runtime consistency~\cite{gannon_cloudnativeapps_2017}. 
Meanwhile, orchestration systems like Kubernetes\cite{Burns_BorgKubernetes_2016} take care of the surrounding tasks, including resource scheduling, lifecycle management, scaling, and request routing \cite{Khan_OrchestrationPlatforms_2017,rodriguez_containerorchestration_2018}.
Together, these technologies have made large-scale service deployments more efficient and practical \cite{Medel_k8sOverhead_2018,8790136}. %

However, containers also come with well-known limitations, including large image sizes, start-up latency, and orchestration overhead. Earlier studies have, for example, measured overheads of up to 10\% in network-heavy workloads~\cite{ref22_morabito2015, KOZHIRBAYEV2017175}.
Recently, WebAssembly has emerged as a new isolation and deployment technology.
It features a lightweight runtime with strong isolation, small binaries and fast startup~\cite{Liu_WASMforContainers_2025,hasselt_WASMEnergy_2022}.
Already many studies have demonstrated Wasm's suitability for serverless and edge scenarios, where container limitations are most pronounced~\cite{wang_EdgeWasm_2021,kjorveziroski_WASMNext_2023,Shillaker_faasm_2020,Kang_HybridWasm4FaaS_WOSC_2025}. 

However, research on WebAssembly beyond serverless and edge use cases remains rare. In the past, the technology was admittedly very limited and unsuitable for cloud deployments. But thanks to WASI and other ecosystem improvements, Wasm is maturing very rapidly. Two recent papers have already examined Wasm as a potential replacement for containers~\cite{Liu_WASMforContainers_2025,kakati_WASMvsContainer_2025}, with mixed results. We argue that more empirical studies are needed. Existing comparisons tend to focus on start-up latency, leaving operator-side questions of resource efficiency and lifecycle management underexplored.  
Plus, in such a fast-moving ecosystem, results showing WebAssembly trailing containers may not hold for long.

In this work, we explore WebAssembly-based service orchestration as a foundation for more efficient cloud deployments. We characterize different workloads for cloud-native environments, investigate which types of workloads stand to benefit most from this approach, and conduct experiments with a particular focus on resource efficiency and service lifecycle management. 

In summary, we make the following contributions:
\begin{itemize}
\item We design and implement \name{}, a proof-of-concept orchestrator for Wasm-based services, addressing scheduling, lifecycle management, and request routing
\item We evaluate our system across a representative set of cloud-native workloads, comparing it against a container-based baseline
\item We discuss the applicability of Wasm-based services based on results from above.
\end{itemize}

The paper is structured as follows: First, we review related work in~\Cref{sec:rw}. From this we extract requirements for \name{} in~\Cref{sec:tool}. ~\Cref{sec:eval}  evaluates \name{} and presents results for resource efficiency, start-up latency, and lifecycle management. Finally, we discuss results in ~\Cref{sec:diss} before concluding in~\Cref{sec:concl}.

\section{Related Work}\label{sec:rw}
Cloud service architectures are built for horizontal scalability and rely heavily on containerization and complex orchestrators like Kubernetes~\cite{ref16_lopez2017} to manage resources at scale.
While containers provide necessary isolation, they introduce some overhead, particularly during I/O operations, context switching, and cold-start events~\cite{ref22_morabito2015, KOZHIRBAYEV2017175}.
These overheads, which can be substantial for resource-constrained or high-density environments, limit the overall efficiency and performance of cloud-native deployments~\cite{ref24_warade2023}.

WebAssembly (Wasm) has emerged as a promising alternative isolation technology~\cite{wang_EdgeWasm_2021}. 
Unlike traditional virtualization or container runtimes, Wasm operates on a lightweight, capability-based model\cite{kakati_WASMvsContainer_2025}. 
This design enables fine-grained resource access control through explicit capabilities, implemented entirely in userspace.
This approach avoids the expensive context switches associated with system calls in container systems, offering a safer and lower-overhead execution environment\cite{Liu_WASMforContainers_2025,Hall_HybridWasm4Edge_MIDDLEWARE_2025}.

These savings should translate into measurable performance benefits, such as lower start-up latency and energy consumption. To test this, we developed a small proof-of-concept orchestrator for WebAssembly services to serve as a benchmarking harness.

We chose to build a minimal system in Rust rather than extend an existing open-source orchestrator. Other existing Wasm orchestrators were either purpose-built for a different serverless/edge environment \cite{Nurul_WASM_IC2E_2021,hasselt_WASMEnergy_2022,Kakati_WASMperformanceCloudContinuum_2024,kjorveziroski_WASMNext_2023} or, in the case of Kubernetes, are so complex that they obfuscate the core orchestration primitives we wanted to evaluate. Our intent was not to build a competitive orchestrator, but to construct a controlled environment in which the essential primitives such as resource scheduling, lifecycle management, scaling, load balancing, and routing could be measured in isolation.

\section{Kirin}\label{sec:tool}
In this paper, we develop \name{}, a WebAssembly-based service orchestrator that can be used to deploy Wasm-based services on a host.
For this, we first lay out the basic requirements that a WebAssembly-based service orchestrator needs. Then, we describe the Architecture of \name{}, how it is used and how the orchestrator works in detail.

\subsection{Requirements}
\paragraph{Lifecycle management} A primary task of an orchestrator is to manage the lifecycle of services. This means that the orchestrator must accept a definition of a service and manage its state throughout its lifecycle. The lifecycle of a service encompasses all its state transitions from service creation to service destruction. Typically, the operations required include \textbf{creating, starting, stopping, updating, and destroying} the service. Consistent state management is required because services may fail, and the orchestrator is responsible for moving the service to an appropriate state from which it can be restarted. 

\paragraph{Resource management} An orchestrator needs to distribute the resources of the host machine across the services that are orchestrated on that machine. This means that it must provide facilities to partition and isolate host resources, such as the CPU and memory. More advanced orchestrators might also limit or partition the throughput on resources such as the network or the file system\cite{rodriguez_containerorchestration_2018}. 

\paragraph{Scaling} Beyond resource management, the orchestrator must allow services to scale based on the load they receive. This means automatically duplicating services to meet increased demand and reducing the number of service instances when demand subsides. When scaling, the orchestrator must account for the host machine's resource limits, thereby connecting these two requirements. Scaling typically entails the automatic scheduling of resources and services, as well as distributing incoming load across service replicas according to a load-balancing policy. Such scaling and load-balancing also enables graceful rollouts and updates, critical for continuous service operation.

\paragraph{Service discovery} Most orchestrators provide a way for each service to discover other services, which is essential for creating distributed applications where components need to communicate with each other. In the dynamic environments typical for microservice systems, services are frequently deployed, scaled, moved, or replaced across infrastructure, which makes static configuration impractical. 

Both Docker~(containerd) and Kubernetes already support these requirements. 
For Wasm-based services, there are existing Runc-compatible shims that support lifecycle and resource management.
However, these projects are still in the incubation stage. The goal of these projects is to integrate Wasm into existing ecosystems, but it remains unclear whether the same abstractions and separation of concerns that work well for Docker (containerd) also make sense for a Wasm runtime.
For example, it is not obvious whether request handling should be delegated to processes running within a container, or whether request-based isolation should instead be provided directly by the runtime.
In this paper, we focus on the core orchestration primitives required to run Wasm-based services. Therefore, we opted to implement a standalone orchestration that natively supports Wasm-based services, which we call \name{}~\footnote{\name{} is available at \url{https://github.com/nymphbox/kirin}}. 

This approach gives us better control over the relevant aspects of serving and managing Wasm services and avoids obscuring performance and lifecycle differences through additional abstraction layers such as runc and Kubernetes.

\subsection{Architecture}
\name{} is a small orchestrator for WebAssembly services implemented in Rust.
We chose Rust because it performs well on energy efficiency benchmarks and facilitates concurrent and multi-threaded programming. Additionally, Rust provides a robust ecosystem of libraries for interacting with WebAssembly, given that the Wasmtime\footnote{\url{https://wasmtime.dev/}} WebAssembly runtime is also implemented in Rust, and Rust generally treats WebAssembly as a first-class citizen in its tooling.

\begin{figure}[htbp]
\centering
\includegraphics[width=\columnwidth]{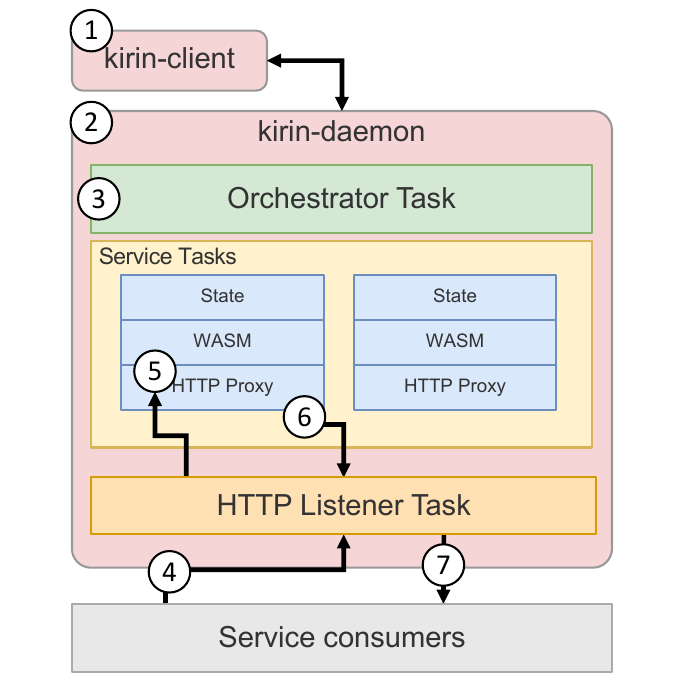}
\caption{\name{} architecture}
\label{fig:recovered-3}
\end{figure}

\name{} (see~\cref{fig:recovered-3}) follows a client-server architecture inspired by Kubernetes, a client that wraps an HTTP-API \circled{1} to interact with the orchestrator daemon \circled{2}. The daemon turns API requests into specific orchestration tasks \circled{3}. The orchestrator owns all running tasks (services), shared state and available service listeners for a single node.
Listeners \circled{4} accept http/grpc connections from consumers and request a service from the orchestrator to spawn service tasks to handle requests. Each service task \circled{5} requests the Wasm function to handle the request and capture the output \circled{6}. The captured output is then handed back to the listener \circled{7} to complete the typical request flow. 

We build \name{} with concurrency and internal asynchronous communication in mind. We leverage the Rust-based Tokio~\footnote{\url{https://tokio.rs/}} runtime to run all orchestration, service and communication tasks, akin to goroutines or lightweight threads in other environments. Through this, we are able to have a single daemon interact with a very large number of http clients and serve Wasm tasks to them.
The \name{} daemon is an HTTP API server built on top of Axum\footnote{\url{https://docs.rs/axum/latest/axum/}}. The http-based API allows a potential multi-node orchestration as we can communicate with a \name{} daemon across a network. 
Internally, all tasks communicate using channels without any other synchronization mechanism.

We use the Wasmtime\footnote{\url{https://bytecodealliance.org/}} runtime to compile and execute WebAssembly components and modules. Wasmtime performs relatively well in benchmarks \cite{kjorveziroski_WASMNext_2023,hasselt_WASMEnergy_2022}. 
For compilation Wasmtime uses Cranelift\footnote{\url{https://cranelift.dev/}} , which is also developed by the ByteCode Alliance. Xu et al. \cite{xu_CopyPatch_2021} show that while Cranelift produces native code that is marginally slower than V8 (2\%), and relatively slower than LLVM (14\%), it compiles code around 10x faster than LLVM. Wang et al. \cite{wang_EdgeWasm_2021} show that Wasmtime and Cranelift outperform other WebAssembly code generators on real-world benchmark tasks. They also show that Wasmtime and Cranelift have the best startup performance, echoing the significantly higher compilation speed of Cranelift.

\subsection{Usage}

Taking inspiration from Kubernetes, \name{} distinguishes between a service specification and a service instance. A service specification is the abstract definition of a service consisting of a name, the associated WebAssembly bytecode, and configuration parameters, such as the port number or the maximum usable memory. 
A service is implemented through a listener task that can spawn service tasks. The listener task is spawned when \name{} receives a start command for a particular service. The listener task accepts TCP connections on the port defined in the service specification and will forward any valid HTTP requests received on that connection to a service task it spawns in response to each request. Thus \name{} provides per-request isolation, which is a stronger variant of isolation than what is provided by a container-based system. However, this creates potential start-up cost per request, as we have to create an isolated environment per request. Each service task instantiates a store. A store is a collection of WebAssembly instances and state and is provided by Wasmtime as the container for all relevant WebAssembly objects such as functions, instances, memories and so on. Stores are designed to live no longer than the WebAssembly instances they hold, and so \name{} follows this pattern by tying these lifetimes directly together, i.e. the store is deallocated as soon as the request is handled, whereas the listener task is designed to be long-running and will exit once the daemon receives the corresponding command. 

\subsection{Orchestration}

The system needs to be able to isolate and partition the host's resources like the file system, the network, the CPU, and memory. 
By default, it is impossible for a WebAssembly module to perform anything other than pure computation, manipulation of its own linear memory, and calling declared and bound host functions. This gives us strong isolation by default. 
To enable more useful programs, the host (\name{}) can define standardized WASI host functions, that allow for additional capabilities such as network and disk access. These capabilities can be enabled on a per-service basis, and are part of the service description in \name{}. 
Besides enabling general capabilities, an orchestrator also needs to manage memory and CPU resource allocation.
For memory, Wasmtime already comes with easy means to define a memory limit that will interrupt and terminate the execution of the service if it exceeds this limit. 
For CPU usage, Wasmtime does not provide a built-in mechanism to limit CPU access but it allows for fine granular instruction counting through fuel or by implementing an epoch time-based interruption mechanism to manage execution time. In \name{}, CPU limiting is thus implemented on top of this epoch mechanism. Users can configure a tick interval at which running Wasm code will be interrupted. Additionally, the starting timestamp of each Wasm service is tracked, so we can determine the actual time that has passed since the start of execution. These two mechanisms taken together implement very simple cooperative timeslicing, which we can use to limit execution of a particular service.

\name{} expects users to transmit the compiled Wasm binaries when creating a new service. The files are stored on disk. In comparison to docker images, such binaries are typically very small, e.g., below 1 MB, and thus relatively cheap to store and send through the network. 
Due to this, startup costs are also low, as the time to start a service is limited only by the file size. The Tokio processes are already running, thus, loading in a new task adds very little overhead.

\section{Evaluation}\label{sec:eval}
In the following, we evaluate whether Wasm-based services show improved performance, resource, and energy efficiency by comparing \name{} with Docker. 
\subsection{Experiment Protocol}
All experiments were run on an \textit{Intel Xeon E3-1230 V2} with 32 GB Memory. We disable Turboboost and set the CPU governor to performance (fixed frequency) for less variation in power measurements.
As a system under test (SUT) we compare \name{} and Docker-CE, running the same services (see \Cref{tab:workloads}) compiled from Rust to binaries or Wasm-binaries. For Wasm isolation, the Rust code is scaffolded with a simple WASI http handler. For container isolation, the Rust code is scaffolded with a simple asynchronous http request handler. Both scaffolds are kept as simple as possible, using the same asynchronous framework. Each workload therefore conceptually represents a service that receives an http request and performs some computation in response. All workloads were deployed with the same resource limits.
We compare against Docker rather than Kubernetes deliberately. 
Our aim is to isolate the core orchestration primitives, and a full cluster orchestrator adds scheduling, networking, and control-plane layers that would add noise to the per-operation measurements.

\begin{table}[]
    \centering
     \caption{Overview of the different services used in the evaluation.}
    \label{tab:workloads}
    \resizebox{\columnwidth}{!}{
    \renewcommand{\arraystretch}{1.5} 
    \begin{tabular}{lp{0.15\columnwidth}p{0.7\columnwidth}}
        Workload & Category & Description \\ \hline
        \textit{hello-world} & External Network & Most basic http request-response loop. \\
        \textit{read-file} & Disk &  Sequential and random reads and writes on a temporary file .\\
        \textit{http-request} & Internal Network & Calls other services in the network using http. \\
        \textit{n-body} & CPU &  N-body problem approximation causing many floating point operations.\\
        \textit{binary-tree} & Memory &  Construction and traversing on a large binary tree in memory. \\
        \textit{matrix-multiplication} & CPU/Memory &  Matrix multiplication with varying matrix sizes.\\ \hline
    \end{tabular}
    }
\end{table}

\paragraph{\textbf{Measurements}}
The aim of this evaluation is to find performance/energy trade-offs in each SUT. 

For service performance, we mainly focus on \textbf{latency} and \textbf{throughput} of the http API of the deployed tasks. 
As part of the evaluations, we also investigate how the throughput metric changes with more deployed instances ($a$). For this, we compute a \textbf{concurrency factor} by comparing the throughput achieved at one concurrency level $a$ to the next $a+1$. 
This gives additional insight into how well each SUT is able to use available resources in parallel. 

For the CPU and memory efficiency metrics, we record the \textbf{CPU utilization} and \textbf{memory utilization} as percentages of available capacity. 
These utilization metrics can then be compared against the number of operations achieved per benchmark, which leads to an efficiency metric that gives a utilization per operation performed. 

For energy consumption, we opted to apply a power modeling approach~\cite{bertran_powerModels_2010}. 
For this, we collected and observed performance counters using perf under stress scenarios while also collecting RAPL\cite{khan_rapl_2018} values.
Through a principal component analysis, we selected \textit{cycles} and \textit{stalls}, which together explain 95\% of the variance in the observed counter data.
Intuitively, cycles and stalls reflect two distinct aspects of power consumption, where the former measures the power consumption when the CPU is actively executing instructions, and the latter measures power consumption in a less energy-intensive state. 
The resulting model with fitted coefficients for the \textit{Intel Xeon E3-1230 V2} is:
\begin{equation}
P = 15.47 + 3.34x_1 + 2.23x_2 + \epsilon
\end{equation}
where $P$ is power in watts, $x_1$ is cycles, $x_2$ is stalled cycles, $\epsilon$ is an error term and the intercept models the average baseline power. The regression achieves an in-sample R2 value of 0.91, which is in line with the literature for similar models. 
The main benefit of using this model over RAPL measurements for our experiments is the sampling frequency difference between RAPL and perf, allowing us a more fine-grained power consumption measurements for short-running workloads.

\paragraph{\textbf{Benchmarks}}

In total, we designed three benchmarks to evaluate the SUT.
First, a \textbf{HTTP Load Test} for individual services. We measure the request latency as the time it takes for a request to be processed and the request throughput as the number of requests per second. We further measure the data transfer rate for throughput. We also measure the success rate across requests. This benchmark is designed to capture how the isolation layer affects each metric across different services. The benchmark can be configured to run concurrently or sequentially and we report results for both configurations. Additionally, varying the concurrency allows us to examine the degree of parallel resource efficiency of each isolation technology. 

Second, a \textbf{Cold-Start Test}, where we create many instances of a service and measure the distribution of cold-start latency, i.e. the time it takes from instance creation to a healthy service ready to take requests. 

Lastly, an \textbf{Orchestration Test}, which randomly selects from a set of plausible orchestration actions to investigate lifecycle management. These actions are: (a) creating an instance, (b) deleting an instance, (c) stopping an instance, (d) updating an instance configuration and (e) listing the current status of all service instances. This benchmark is designed to model the regular operation of an orchestration system. During the orchestration test we configure the http-request service to make requests to other services at random to model inter-service communication.

\subsection{Results}
In the following we present the results of the three described benchmarks, comparing energy efficiency, resource efficiency, throughput and latency between \name{} and Docker. 

\subsubsection{HTTP Load Test}
\begin{table*}[h]
\centering
\caption{HTTP benchmark: \name{} relative to docker baseline, expressed as
$\Delta = (\textrm{\name{}} - \textrm{docker})/\textrm{docker}$}
\label{tab:http-benchmark-results}
\resizebox{\linewidth}{!}{%
\begin{tabular}{lcccccc}
\toprule
& \multicolumn{3}{c}{Efficiency} & \multicolumn{3}{c}{Performance} \\
\cmidrule(lr){2-4}\cmidrule(lr){5-7}
Workload
& \shortstack{$\Delta$ Energy eff.\ $\uparrow$\\\scriptsize[\%]}
& \shortstack{$\Delta$ CPU eff.\ $\uparrow$\\\scriptsize[\%]}
& \shortstack{$\Delta$ Mem.\ eff.\ $\uparrow$\\\scriptsize[\%]}
& \shortstack{$\Delta$ Throughput $\uparrow$\\\scriptsize[\%]}
& \shortstack{$\Delta$ Mean lat.\ $\downarrow$\\\scriptsize[\%]}
& \shortstack{$\Delta$ Tail lat.\ $\downarrow$\\\scriptsize[\%]} \\
\midrule
\multicolumn{7}{l}{\textit{I/O workloads}} \\
\midrule
hello-world           & \textcolor{lgreen}{+186.0} & \textcolor{lgreen}{+207.0} & \textcolor{lgreen}{+221.0} & \textcolor{lgreen}{+213.0} & \textcolor{lgreen}{-60.0}  & \textcolor{lgreen}{-68.0}  \\
read-file             & \textcolor{lgreen}{+63.0}  & \textcolor{lgreen}{+58.0}  &\textcolor{lgreen}{ +98.0 } & \textcolor{lgreen}{+95.0}  & \textcolor{lgreen}{-54.0}  & \textcolor{lgreen}{-86.0}  \\
http-request          & \textcolor{lgreen}{+3.0}   & \textcolor{red!60}{-7.0 }  & \textcolor{lgreen}{+2.9}   & \textcolor{lgreen}{+6.7}   & \textcolor{lgreen}{-4.0 }  & \textcolor{red!60}{+1.0}   \\
\midrule
\multicolumn{7}{l}{\textit{CPU, memory and mixed workloads}} \\
\midrule
n-body                & \textcolor{red!60}{-42.0 } & \textcolor{red!60}{-44.0}  & \textcolor{red!60}{-34.0}  & \textcolor{red!60}{-36.0}  & \textcolor{red!60}{+67.0}  & \textcolor{lgreen}{-3.0}   \\
binary-tree           & \textcolor{lgreen}{+2.86 } & \textcolor{red!60}{-2.0}   & \textcolor{red!60}{-30.0}  & \textcolor{red!60}{-31.7}  & \textcolor{red!60}{+47.0}  & \textcolor{red!60}{+39.0}  \\
matrix-multiplication & \textcolor{red!60}{-76.0}  & \textcolor{red!60}{-74.1 } & \textcolor{red!60}{-72.0}  & \textcolor{red!60}{-72.0}  & \textcolor{red!60}{+255.0} & \textcolor{red!60}{+163.0} \\
\bottomrule
\end{tabular}%
}

\end{table*}

\Cref{tab:http-benchmark-results} shows the main results of the http Load test. Here, we can see a strong divide between I/O bound workloads and memory/CPU bound workloads, where \name{} dominates I/O workloads but underperforms for CPU/Memory bound workloads.
Hence, we can see that WebAssembly is generally preferable for short-lived and I/O-bound services, where it outperforms Docker in every metric. We see that the less intensive a particular workload is, the more the isolation overhead of Docker drags on performance and efficiency. Conversely, Docker clearly outperforms for compute-intensive tasks, and this advantage seems to increase the more CPU-intensive a task is, with the matrix workload taking the longest time and simultaneously offering the worst results for WebAssembly. This confirms that WebAssembly is not the isolation technology of choice for orchestrating long-running, CPU-intensive workloads.

We also run the http benchmark with the hello-world workload in different concurrency configurations varying the number of threads and connections to evaluate how well each technology utilizes resources in parallel. We avoid excessive contention on the test machine by limiting the number of threads and connections to less than the available cores on the machine. We choose the hello-world benchmark because it is the simplest workload that reveals the basic differences in both technologies. \Cref{tab:http-concurrency} shows that WebAssembly provides a much higher base throughput in all concurrency settings, and has better incremental scaling factors and a better overall concurrency factor (3.4x for Wasm vs. 2.8x for Docker). Note that this is despite \name{} implementing an instance-per-request scheme versus the single Docker instance tested.

\begin{table}[htbp]
\centering
\caption{Requests per second and concurrency factors for different concurrency levels in the http benchmark. (Factors are relative to prior level).}
\label{tab:http-concurrency}
\resizebox{\columnwidth}{!}{
\renewcommand{\arraystretch}{1.5} 
\begin{tabular}{l|cccc}
\hline
Concurrency Level & 1       & 2                       & 4                       & 6                       \\ \hline
\name{}           & $11469$ & $13671$ ($1.2\text{x}$) & $29206$ ($2.1\text{x}$) & $38261$ ($1.3\text{x}$) \\
Docker            & $4635$  & $7559$ ($1.6\text{x}$)  & $11169$ ($1.4\text{x}$) & $12981$ ($1.1\text{x}$) \\ \hline
\end{tabular}%
}
\end{table}

\subsubsection{Cold-Start Test}

For the cold-start benchmark, we try to spawn as many instances of a service with a particular workload as possible for 60 seconds without putting the system under load.

Given that the results are (naturally) extremely similar across workloads, we only report the results for the hello-world workload in \cref{fig:cold_start_perf}. For this benchmark, the cold start latency is defined as the time it takes to start a new service instance with a particular workload from scratch until it is ready to respond to requests. 
Note that due to the way we execute the benchmarks via our runner, this time includes process creation and other initialization work, which however is the same for both technologies and thus will not affect the comparison.

\begin{table}[]
    \centering
    \caption{Performance metrics of the cold start benchmarks for the hello-world workload.}
    \label{fig:cold_start_perf}%
    \renewcommand{\arraystretch}{1.5} 
    \resizebox{\columnwidth}{!}{%
\begin{tabular}{lrrrrrl}
\toprule & \multicolumn{3}{c}{Efficiency} & \multicolumn{3}{c}{Performance} \\
\cmidrule(lr){2-4}\cmidrule(lr){5-7}
Runtime & \shortstack{Energy eff.\ $\uparrow$\\\scriptsize[instances/W]}
& \shortstack{CPU eff.\ $\uparrow$\\\scriptsize[req./\% cpu util.]}
& \shortstack{Mem.\ eff.\ $\uparrow$\\\scriptsize[req./\% mem]}
& \shortstack{Throughput $\uparrow$\\\scriptsize[instances/s]}
& \shortstack{Mean lat.\ $\downarrow$\\\scriptsize[s]}
& \shortstack{Tail lat.\ $\downarrow$\\\scriptsize[s]} \\
\midrule
kirin & \textbf{8.91} & \textbf{10.16} & \textbf{0.0020} & \textbf{12.02} & \textbf{0.0850} & \textbf{0.0900} \\
Docker & 5.50 & 7.42 & 0.0017 & 3.06 & 0.3387 & 0.4542 \\
\bottomrule
\end{tabular}%
}

\end{table}

The efficiency metric for this benchmark is the number of created instances divided by the total power estimated by the energy model, and thus very similar to the efficiency metric in the http Load Test. The throughput metric for this benchmark is correspondingly the number of created instances per second. \Cref{fig:cold_start_perf} shows that WebAssembly instances via \name{} are far cheaper to initialize, and that we therefore can create many more instances than Docker containers, as the throughput metrics show. We are able to create around 720 service instances with \name{}, but only 180 instances with Docker, as the four times higher throughput indicates. The tail latency, which shows how long instance creation can take in the worst case, is five times better for WebAssembly, and very close to the mean cold start latency. These performance characteristics are reflected in the efficiency metrics. As we have speculated, the driver of the better initialization performance is significantly smaller WebAssembly modules that have less overhead due to a more static isolation. Thus WebAssembly uses the available memory and the available CPU cores more efficiently than Docker, resulting in a 1.62x higher energy efficiency. 

\begin{figure}
\centering
  \centering
  \includegraphics[width=\linewidth]{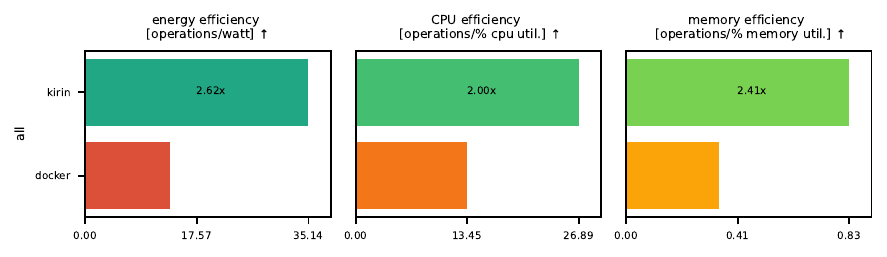}
  \includegraphics[width=\linewidth]{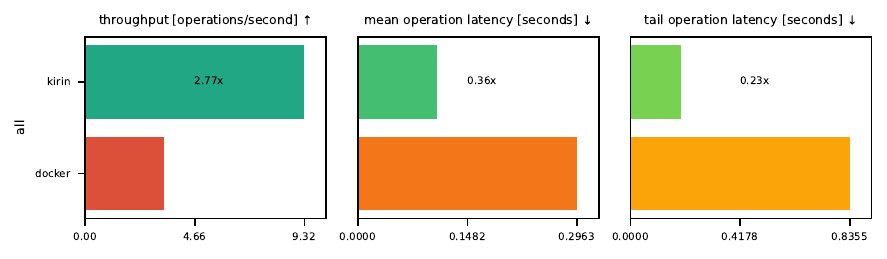}
\caption{Performance and Efficiency of Orchestration Benchmark}
    \label{tab:orchestraion_results}
\end{figure}

\subsubsection{Orchestration Test}

\begin{figure}
    \centering
    \includegraphics[width=\linewidth,trim={0 0 3.4cm  0},clip]{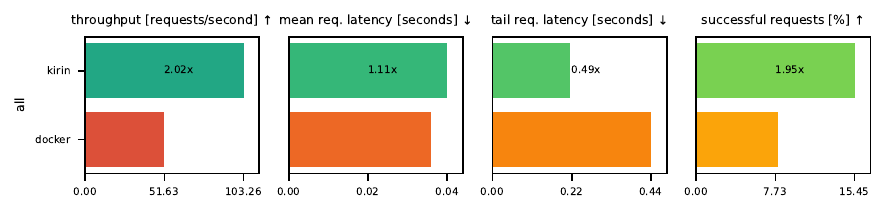}
    \caption{HTTP performance metrics in the orchestration benchmark}
    \label{fig:http-orchestration}
\end{figure}

For the orchestration benchmark, we selected a random plausible orchestration action, e.g., creating, deleting, and updating a service, ensuring that we only selected actions that were possible given the systems state. For the update action, we have avoided modifying the base image for the service, as this would be so costly for Docker (requiring a full rebuild of the container) that it would make Docker wholly uncompetitive, leading to uninteresting results. Beyond these orchestration actions, we put each created service under load similar to what we did in the HTTP benchmark. This allows us to take into consideration how disruptive the orchestration actions are to the continued functioning of the services, and how http throughput and latency are affected by each orchestrator. 
Starting with the performance metrics for the orchestration actions, \cref{tab:orchestraion_results} shows operation throughput as well as mean and tail latency across all operations. Evidently, \name{} via WebAssembly is able to outperform Docker in terms of pure orchestration performance across all metrics. This is primarily driven by the lighter isolation model, where we avoid additional work by the operating system, leading to better startup and shutdown performance, which dominates the benchmark results. 
The higher number and faster startup are also reflected in an overall better http performance (see~\cref{fig:http-orchestration}), where \name{} could service many more successful requests. 
One reason for the increased throughput is that the WebAssembly services were much faster in serving requests when started through random orchestration events, hence allowing overall more requests to succeed over time.

\section{Discussion}\label{sec:diss}
In the following we analyze the results of the evaluation in~\cref{sec:eval} and review how these results motivate a broader adoption 
of WebAssembly for cloud native architectures. 

\subsection{Result Analysis}
Summarizing the results, we conclude that \name{} provides more lightweight lifecycle management, primarily driven by the improved startup and shutdown characteristics of WebAssembly, as we also observed in the cold-start benchmarks. 
Recall that this result is expected because WebAssembly modules have a smaller footprint than Docker images and require less work to be performed at initialization compared to containers.
Conversely, this leads to better shutdown performance as there are fewer structures to release and clean up than with containers. WebAssembly thus offers far better operation throughput, which is also reflected in the improved efficiency metrics compared to Docker containers, where the higher throughput explains much of the efficiency gains.
Moreover, because there is less disruption when performing orchestration operations while the system is under load, the HTTP metrics also tend to be better for WebAssembly when compared to Docker, with the exception of the mean latency.
These results are also in line with the observations from the cold-start benchmarks, which show overall better efficiency when starting new service instances.
Beyond the improvement in lifecycle management and cold start, we confirmed that Wasm performs well for I/O-bound workloads but underperformes for long running CPU or Memory based workloads. 
These trends also show overall improved efficiency at higher concurrency factors.

\subsection{A case for WebAssembly in cloud native architectures}
\begin{figure}
    \centering
    \includegraphics[width=\columnwidth]{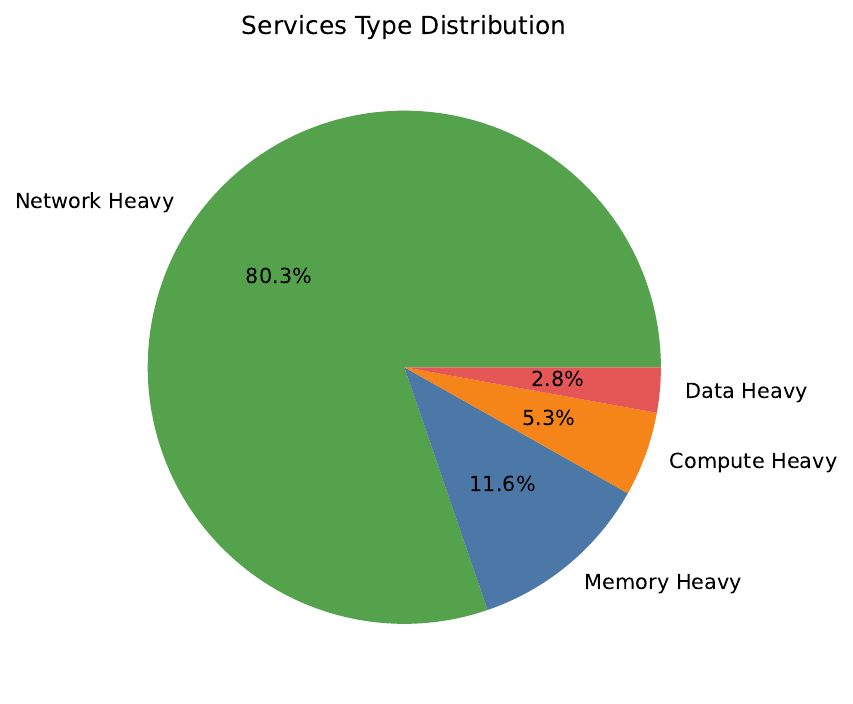}
    \caption{Dominant resource need of service in the Alibaba Cluster Trace in one day~\cite{luo2022Prediction}}
    \label{fig:service-overview}
\end{figure}
Coupled with the good performance on I/O dominant tasks, we can clearly see a space where WebAssembly-based web services make sense in a larger cloud-native architecture.
Moreover, we do not claim that WebAssembly matches native execution speed; the matrix and n-body results show clearly that it does not.
Indeed, we argue that most web services are not bound by CPU or memory but by network and IO, so the relevant comparison is isolation cost, not peak compute. See~\cref{fig:service-overview} as an example. 
Here, we took one day's worth of tracing data from the Alibaba Cluster Trace 2022 dataset~\cite{luo2022Prediction} and reviewed what the most dominant resource was per recorded service during its lifetime\footnote{We arrived at this distribution by classifying all microservices in the data into each dominant-usage category (highest accumulated usage). Each service is first reduced to a service-level feature profile, its Compute, Memory, Network, and Database intensities are standardized across services, and the service is then assigned to the category corresponding to its maximal standardized intensity.}.
We observe that networked operations (HTTP, gRPC, events) dominate the distribution.
Here, WebAssembly could be a feasible alternative to containers, offering smaller footprints and higher energy efficiency.
This also confirms the trend for WebAssembly-based serverless platforms\cite{Shillaker_faasm_2020,Kang_HybridWasm4FaaS_WOSC_2025}, which benefit the most from cold-start and orchestration overhead reduction because they typically host short-lived networked tasks.
However, we still need to integrate this technology better into already existing orchestrators such as Kubernetes, to allow for such hybrid systems that combine the strengths of both.

\section{Conclusion}\label{sec:concl}
In this work, we examined WebAssembly as an isolation technology for cloud-native services, with a particular focus on resource efficiency and service lifecycle management.
For this, we designed, implemented, and evaluated \name{}. We compared WebAssembly-based web services against Docker.
Here, we have shown that WebAssembly offers substantial advantages for I/O-bound services, where its lightweight isolation model translates directly into higher throughput, lower tail latency, and markedly better energy efficiency. 
Most notably, \name{} achieves a 2.62x improvement in energy efficiency over Docker under continuous orchestration load, sustains roughly four times the instance creation throughput, and improves cold-start energy efficiency by 1.62x.
At the same time, our results delineate the boundaries of this approach. For long-running, CPU- or memory-bound workloads, the overhead of WebAssembly's execution model outweighs the gains from cheaper isolation, and Docker remains the more appropriate choice. 
WebAssembly is therefore best understood not as a wholesale replacement for containers, but as a complementary isolation technology.

We argue that most web services are typically not performing heavy computation or need vast amounts of memory, therefore, inviting the WebAssembly benefits to cloud native services beyond its adoption for serverless and edge platforms.
However, our evaluation was deliberately scoped to isolated workloads and a purpose-built orchestrator, which leaves open the question of how these results carry over to production environments. For future work, we aim to examine WebAssembly-based services in the context of large microservice topologies and in the context of mature and distributed orchestrators such as Kubernetes.

\bibliographystyle{ieeetr}
\bibliography{refs}

\end{document}